\documentclass[prd,twocolumn,aps,preprintnumbers,amsmath,amssymb,nofootinbib]{revtex4}

\usepackage{graphicx}
\usepackage{dcolumn}
\usepackage{bm}
\usepackage{amssymb}
\usepackage{mathrsfs}
\usepackage{amsmath}
\usepackage{epsfig}
\usepackage{color}
\usepackage{hhline}

\allowdisplaybreaks[4]

\begin{document}

\title{Constraints and tests of the OPERA superluminal neutrinos}

\author{Xiao-Jun Bi}
\author{Peng-Fei Yin}
\author{Zhao-Huan Yu}
\author{Qiang Yuan}

\affiliation{Key Laboratory of Particle Astrophysics, Institute of High
Energy Physics, Chinese Academy of Sciences, Beijing 100049,
China}

\begin{abstract}

The superluminal neutrinos detected by OPERA indicates Lorentz
invariance violation (LIV) of the neutrino sector at the order of
$10^{-5}$. We study the implications of the result in this work.
We find that such a large LIV implied by OPERA data will make the
neutrino production process $\pi \to \mu + \nu_\mu$ kinematically
forbidden for neutrino energy greater than about $5$ GeV. The
OPERA detection of neutrinos at $40$ GeV can constrain the LIV
parameter to be smaller than $3\times 10^{-7}$. Furthermore the
neutrino decay in the LIV framework will modify the neutrino
spectrum greatly. The atmospheric neutrino spectrum measured by
IceCube can constrain the LIV parameter to the level of
$10^{-12}$. The future detection of astrophysical neutrinos of
Galactic sources is expected to be able to give even stronger
constraint on the LIV parameter of neutrinos.

\end{abstract}

\maketitle

{\it Introduction}---Recently the OPERA collaboration reported an 
astonishing discovery that the neutrinos from CERN to the Gran Sasso 
Laboratory may be superluminal \cite{opera}. The arrival time of the 
muon neutrinos with average energy $\sim17$ GeV is earlier by
\begin{equation}
\delta t = (60.7 \pm 6.9(\text{stat.}) \pm 7.4(\text{sys.}))\,\text{ns}
\end{equation}
than that of light, which corresponds to a difference of the muon neutrino
velocity and the speed of light of
\begin{equation}
\delta_{\nu\gamma} = (v_\nu -c)/c = (2.48 \pm 0.28 (\text{stat.})
\pm 0.30 (\text{sys.})) \times 10^{-5} \ ,
\end{equation}
with significance level of $6\sigma$. The energy dependence study of
OPERA shows that the velocity difference $\delta_{\nu\gamma}$ is almost
independent of energy. It was found that, by splitting the events into
two groups with energies below or above 20 GeV, the velocity difference
is given by $\delta_{\nu\gamma} = (2.16 \pm 0.76 \pm 0.30)\times 10^{-5}$
for $\langle E\rangle=13.9$ GeV and $\delta_{\nu\gamma} = (2.74 \pm 0.74
\pm 0.30)\times 10^{-5}$ for $\langle E\rangle=42.9$ GeV.
The results above have been confirmed recently by a test performed 
using a beam with a short-bunch time-structure allowing to measure 
the neutrino time of flight at the single interaction level. The new 
analysis show consistent result with that at the first version.

Actually the earlier experiments also gave
similar results about the velocity of neutrinos, but with lower
significance. In 2007 the MINOS experiment at FERMILAB measured
the neutrino speed for energies around $3$ GeV and found
\cite{minos}
\begin{equation}
\delta_{\nu\gamma} = (5.1\pm 2.9)\times 10^{-5} \ ,
\end{equation}
which seems to be consistent with that by OPERA. Results from even earlier
experiments at FERMILAB showed a similar value of $\delta_{\nu\gamma}$ at
the order of $10^{-5}$ \cite{fermilab}. However, the supernova neutrinos
from SN1987A set a much more stringent limit of $\delta_{\nu\gamma} <
2\times 10^{-9}$ for tens of MeV electron neutrinos \cite{sn1987a}.

Soon after the OPERA result was reported there were many works to discuss
the implications and possible explanations to it \cite{early,superluminal,
Dvali:2011mn,Giudice:2011mm,Cohen:2011hx}. The analysis in these works seem
to get a consensus that the effect should be neutrino flavor independent
so that the coherent neutrino oscillation is not destroyed. Furthermore,
the effect as a function of energy should be a step-like function in
order that it is large and energy independent at OPERA and FERMILAB
energies ($>$GeV), while at low energies it is very small to be consistent
with the SN1987A result.

Phenomenologically such a superluminal neutrino can be easily accommodated
in a Lorentz invariance violation (LIV) model. Coleman and Glashow
(hereafter we refer as CG) constructed such a model where LIV was
expressed as a modified energy-momentum relation
\begin{equation}
E^2=m^2+|\vec{p}|^2+\xi |\vec{p}|^2\ ,
\label{CG}
\end{equation}
assuming that the Lorentz invariance is violated perturbatively in the
context of conventional quantum field theory \cite{Coleman:1998ti}.
In such framework, the neutrino velocity is given by $v_\nu = 1 -
m^2/2|\vec{p}|^2 +\xi/2 $ taking $c=1$. For high energy neutrinos we have
$\delta_{\nu\gamma} = \xi/2 $.

In this work we will assume that the LIV term is flavor independent and
given by the form in Eq. (\ref{CG}) at high energies. At low energy, the
LIV term is suppressed by some specific models, such as that given in
\cite{Dvali:2011mn,Giudice:2011mm}, to be consistent with SN1987A data. 
As studied in the CG paper, many phenomenologies changed significantly. 
Especially for a large $\xi$ of the order ${\cal O}(10^{-5})$, the processes
$\pi^+ \rightarrow \mu^+ + \nu_\mu$ and $\mu^- \rightarrow e^- + \nu_\mu
+\bar{\nu}_e$ are even forbidden so that no high energy neutrinos could
be produced. In the next section we will show the constraints on the LIV
parameters by such processes.

In addition, Eq. (\ref{CG}) can also induce the neutrino radiative
decay $\nu_i \rightarrow \nu_j+\gamma$ and the three body decay
$\nu_i \rightarrow l_i +\nu_e+e$ or $\nu_i \rightarrow \nu_i
+e+e$. Such processes will lead to the fast energy loss of
neutrinos once the threshold is reached. We will discuss the
consequence on the atmospheric and astrophysical neutrino spectra
when considering such processes.

{\it Constraints from neutrino related processes}---
We first consider the process $\pi^+ \rightarrow \mu^++\nu_\mu$.
For $\xi > 0$, the LIV term in Eq. (\ref{CG}) is equivalent to an
``effective'' mass term. If this effective mass is too large, we
will expect kinematics forbiddance of the pion decay. To derive
the threshold condition we get the effective mass of superluminal
neutrino following CG \cite{Coleman:1998ti} as
\begin{equation}
(m_\nu^\mathrm{eff})^2 \equiv m_\nu^2 + \xi |\vec{p}|^2
= m_\nu^2 + 2 \delta_{\nu\gamma}|\vec{p}|^2,
\end{equation}
where we take the LIV parameters of pion and muon to be zero as they
can be much smaller than the neutrino LIV parameter. We will give a
discussion on the LIV parameters of pion and muon in the next section.
Neglecting neutrino mass $m_\nu$ we get the threshold condition for
$\pi^+$ decay $\pi^+ \to \mu^++\nu_\mu $ is $m_\pi \ge m_\mu + 
m_\nu^\mathrm{eff}\approx m_\mu + E_{\nu}\sqrt{\xi/(1+\xi)}$. In the 
laboratory frame we have
\begin{equation}
E_\nu \le (m_\pi - m_\mu) \sqrt{1+1/\xi}\ .
\label{threshold}
\end{equation}
This inequality means that neutrinos generated by pion decay have a
maximum energy beyond which the decay process $\pi^+ \to \mu^++\nu_\mu$
can not occur at all.

\begin{figure}[!htb]
\begin{center}
\includegraphics[width=0.8\columnwidth]{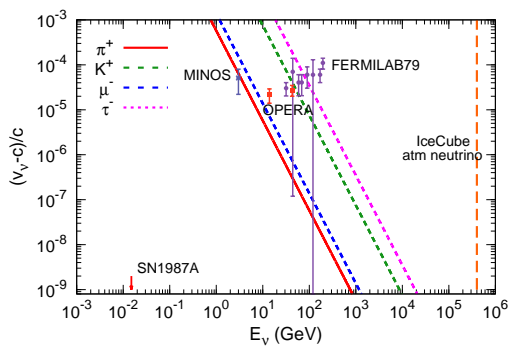}
\caption{The neutrino velocity constrained by the neutrino energy, for
processes $\pi(K)^+\to \mu^+\nu_\mu$, $\mu(\tau)\to \nu_{\mu(\tau)}+
e(\mu)+\bar{\nu}_{e(\mu)}$.}
\label{fig:neutrino}
\end{center}
\end{figure}

Similarly the processes $K^+ \rightarrow \mu^++\nu_\mu$,
$\mu \rightarrow \nu_{\mu} + e + \bar{\nu}_{e}$ and
$\tau \rightarrow \nu_{\tau}+ e(\mu)+\bar{\nu}_{e(\mu)}$
should also be forbidden at high energies.
In Fig. \ref{fig:neutrino} we show the constraints on
the neutrino velocity $(v_\nu-c)/c$ for different neutrino energies
with different processes. The region above the line should be excluded
if we detect the neutrinos with corresponding energies. The experimental
data by FERMILAB79, MINOS and OPERA are also shown. From Fig.
\ref{fig:neutrino} we can see that the process $\pi \rightarrow \mu\nu_\mu$,
through which the OPERA neutrinos are produced, gives the most stringent
constraint. Process $\mu \to \nu_{\mu} + e + \bar{\nu}_{e}$ gives a
similar but a little weaker constraint. It is shown that the detection
of neutrinos with energies as high as $43$ GeV by OPERA can set a
constraint $\delta_{\nu\gamma} < 3\times 10^{-7}$, while the $\sim 200$ GeV
neutrinos detected at Fermilab give $\delta_{\nu\gamma} < 2\times 10^{-8}$.

Actually the highest energy neutrinos detected are the atmospheric neutrinos
at IceCube, which reach $\sim 400$ TeV \cite{icecube}. These neutrinos are
generated by the same process as OPERA neutrinos, i.e. through $pp
\to \pi(K)\to \mu+\nu_\mu$. From Fig. \ref{fig:neutrino} we can see that
the IceCube neutrinos can constrain LIV parameter down to $\sim 4\times
10^{-15}$.

\begin{figure}[!htb]
\begin{center}
\includegraphics[width=0.8\columnwidth]{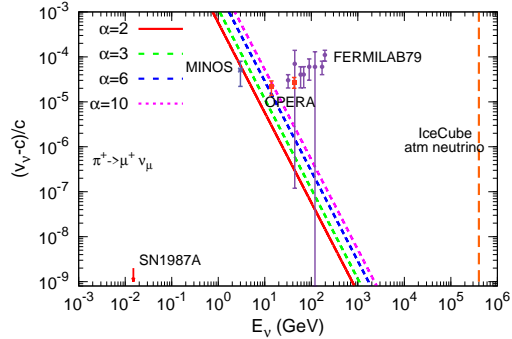}
\caption{The neutrino velocity constrained by the neutrino energy for the
process $\pi^+\to \mu^+\nu_\mu$, for different values of $\alpha$.}
\label{fig:neutrino_alpha}
\end{center}
\end{figure}

In the following we extend the CG model with a more general case
that the dispersion relation is modified as
\begin{equation}
E^2=m^2+|\vec{p}|^2 + \beta |\vec{p}|^\alpha\ .
\label{gene}
\end{equation}
We use the same particle decay processes to constrain the model parameters.
The constraints from neutrino energy for $\pi^+ \to \mu^+ + \nu_\mu$ are
shown in Fig. \ref{fig:neutrino_alpha}, for different values of $\alpha$.
The results show that larger $\alpha$ will give weaker constraints on the
parameters.

{\it The atmospheric and astrophysical neutrino spectra} --- We
have discussed the constraints on the LIV parameter of neutrinos
according to the processes of $\pi\to\mu+\nu_\mu$ and
$\mu\to\nu_\mu+e+\nu_e$. These processes can only constrain the
difference of the LIV parameters between the parent particle and
the daughter particles. Therefore to give stronger constraint on
the neutrino LIV parameter we have to give constraints on $\pi$
and $\mu$ first.

For superluminal $\pi^+$ it will lose energy quickly by radiative emission
$\pi^+\to\pi^+\gamma$. IceCube experiment has measured the atmospheric
neutrino spectrum up to $\sim 400$ TeV, which fits very well with the
model calculation \cite{icecube}. This means that the high energy charged
pions do not lose much energy before they decay to neutrinos. This can give
a strong constraint on the process $\pi^+\to\pi^+\gamma$. The threshold of
this process is $E_\pi > m_\pi/\sqrt{c_\pi^2-c^2}= m_\pi/\sqrt{\xi_\pi}$
\cite{Coleman:1998ti}. This leads to an upper bound of $\xi_\pi < 10^{-14}$
when the maximum energy of pion is about $2$ PeV.

If the maximum attainable velocity of muon is smaller than $c$, it is
still possible to have the process $\pi\to\mu+\nu_\mu$ with superluminal
$\nu_\mu$. In such a case the LIV parameter is constrained by the process
$\gamma\to \mu\mu$. As the $\gamma$ energy as high as $80$ TeV has been
detected at Crab Nebula \cite{Aharonian:2004gb}, we have $|\xi_\mu| <
10^{-11}$. Therefore the LIV parameter of neutrinos can be constrained 
at most to the order $\sim 10^{-11}$ according to pion decay \cite{table}.
More stringent constraint on the flavor dependent LIV parameter was 
set in \cite{hep-ph/9811241}.

In the following we study how to give more stringent constraints
on the neutrino LIV parameter.
For superluminal neutrino with the LIV dispersion
relation Eq. (\ref{CG}), it will lose energy through radiative
decay $\nu_i\to \nu_j+\gamma$ or the three body decay. The three
body decay can be a charged current (CC) process $\nu_\mu \to
\mu+\nu_e+e$ or a neutral current (NC) process $\nu_\mu \to
\nu_\mu+e+e$. Among these three processes, the CC process has the
largest decay rate, but the threshold energy is also the highest.
The radiative process has the lowest decay rate with lowest
threshold energy.
In \cite{Cohen:2011hx} the OPERA neutrino energy loss is studied when
propagating from CERN to Gran Sasso taking $\xi =5\times 10^{-5}$.
As we have shown that neutrinos with energy up to tens of GeV can
not be produced if $\xi$ is so large, we will study the energy
loss process with a much smaller $\xi$. As argued above if muons
are involved, the LIV parameter of neutrinos can only be
constrained to the order $10^{-11}$. Therefore in the following we
focus on the NC process which involves electrons in the final
states, as discussed in \cite{Cohen:2011hx}. The processes $e\to
e\gamma$ and $\gamma\to ee$ give $|\xi_e| < 10^{-15}$ by
observation of the Crab Nebula \cite{Aharonian:2004gb}.

\begin{figure}[!htb]
\begin{center}
\includegraphics[width=0.8\columnwidth]{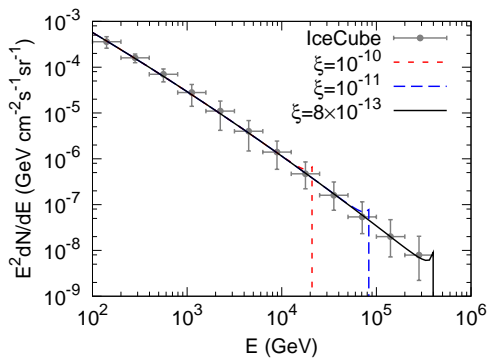}
\caption{The expected atmospheric neutrino spectra with LIV parameters
$\xi=10^{-10}$, $10^{-11}$ and $8\times10^{-13}$ respectively. The measured
data by IceCube is also plotted \cite{icecube}.}
\label{fig:nuspec_atm}
\end{center}
\end{figure}

The calculation of the three body decay rate is very complicated if the
LIV effect of final state particles is taken into account. Working out
the integration of final state phase space, we get
\begin{equation}
\frac{{\rm d}E}{{\rm d}t} = - \langle\Delta E\rangle\Gamma \approx
- 10^{-3} \left( \frac{E}{\text{TeV}}
\right)^6 \left( \frac{\xi}{10^{-10}}\right)^3\text{GeV s}^{-1} \ ,
\label{rate}
\end{equation}
where $\langle\Delta E\rangle$ and $\Gamma$ are the averaged
energy loss of each decay and the decay width. A detailed calculation 
in the laboratory system has been given in \cite{LiM}.

To calculate the neutrino spectra with the energy loss due to decay, we
follow the way of calculating the spectra of ultra high energy (UHE)
protons propagating in the cosmic microwave background (CMB) field
\cite{Berezinsky:2002nc}. The detected neutrino spectra can be written as
\begin{equation}
\phi(E)=\frac{q_0(E_0)}{4\pi d^2}\frac{{\rm d}E_0}{{\rm d}E},
\end{equation}
where $q_0(E_0)$ is the production spectra per unit time at the source,
$d$ is the distance of the source, $E_0$ and $E$ are the energies at
the source and detector respectively. The energy interval dilation
${\rm d}E_0/{\rm d}E$ is
\begin{equation}
\frac{{\rm d}E_0}{{\rm d}E}=\exp\left[\int_0^{t=d/c}\frac{{\rm d}
b[E_0(t')]}{{\rm d}E_0(t')}{\rm d}t'\right],
\end{equation}
where $b(E)\equiv-{\rm d}E/{\rm d}t$ is the energy loss rate in
Eq. (\ref{rate}). Note here we consider the local neutrino sources.
For cosmological sources as discussed in \cite{Berezinsky:2002nc}
Eqs. (9) and (10) are more complicated.

We first study the atmospheric neutrinos. In \cite{icecube} the
up-going neutrinos from $100$ GeV to $400$ TeV were detected by
IceCube. We take a rough travel distance of $10^4$ km for the
neutrinos with zenith angle $124^\circ - 180^\circ$. In Fig.
\ref{fig:nuspec_atm} we show the neutrino spectra with LIV
parameters $\xi = 10^{-10},\, 10^{-11}$, and $8\times 10^{-13}$
respectively. It is shown that for specified distance of the
source, there is a maximum energy of the neutrinos above which the
energy loss is too fast that the neutrinos can not reach us with
high energies. According to the energy loss rate in Eq.
(\ref{rate}), we have $E_{\rm GeV}=(E_{\rm 0,GeV}^{-5}+5\times
10^9\xi^3t_{\rm sec})^{-1/5}$. For $E_0\rightarrow\infty$,
$E_{\rm max}\propto t^{-1/5}=(d/c)^{-1/5}$. For
$\xi=8\times10^{-13}$ we have $E_{\rm max}\approx 400$ TeV for
atmospheric neutrinos. According to the fact that IceCube has
detected atmospheric neutrinos up to $400$ TeV, we can set a bound
of the LIV parameter of neutrino $\xi \lesssim 8\times10^{-13}$.

\begin{figure}[!htb]
\begin{center}
\includegraphics[width=0.8\columnwidth]{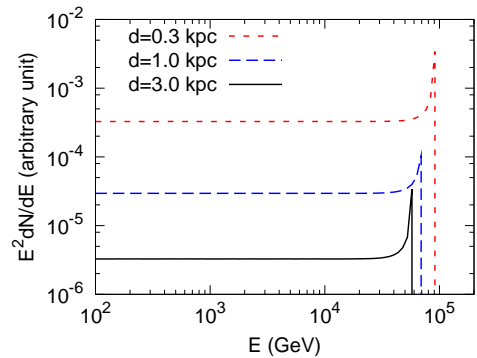}
\caption{Expected neutrino spectra with LIV parameter
$\xi=10^{-15}$, for sources at distances $0.3$, $1.0$ and $3.0$
kpc respectively. The source spectrum is assumed as $E^{-2}$.}
\label{fig:nuspec_snr}
\end{center}
\end{figure}

Then we study the effect of energy loss on the astrophysical
neutrino spectrum. Because these neutrinos will propagate very
long distance before reaching the Earth we expect the spectral
measurement can probe even smaller LIV parameters of neutrinos. Up
to now, no high energy astrophysical neutrinos were detected.
However, as the well believed sources of Galactic cosmic rays
(CRs), supernova remnants (SNRs) are expected to be good
candidates of astrophysical neutrino sources. Fig.
\ref{fig:nuspec_snr} gives an illustration of the astrophysical
neutrino spectra taking into account the LIV effect with much
smaller parameter $\xi=10^{-15}$. Three distances of the sources,
$0.3$, $1.0$ and $3.0$ kpc are adopted. In the calculation the
source spectrum is assumed to be $\propto E^{-2}$. We can see that
for sources with distance $\sim$kpc the cutoff energy is about
several tens to $100$ TeV for this small LIV parameter. The
$\gamma$-ray observations of the Galactic SNRs show that the
accelerated energies of particles can well exceed TeV, which
enables the production of neutrinos with energies much higher than
TeV. Therefore the observation of cutoff at neutrino spectra in
the future can set a very strong constraints on the LIV parameter.

Finally the Greisen-Zatsepin-Kuzmin (GZK) neutrinos are expected
when the ultra-high energy cosmic rays (UHECRs) interact with the
CMB photons and generate the GZK cutoff of UHECRs
\cite{Yoshida:1993pt}. It is expected that the IceCube might
discover such neutrinos in the near future. However, as shown
above the decay mean free path for tens TeV neutrinos is only of
kpc scale. We might have no chance to detect any neutrinos from
extragalactic sources if they are superluminal. 
LIV at the proton sector may also suppress the GZK
neutrinos \cite{Scully:2010iv}.


{\it Conclusion}---We studied the implications of the claim of
superluminal neutrinos by the OPERA collaboration. The observation
generally indicates LIV in the neutrino sector. In such a
framework, we find the neutrino production process
$\pi^+\to\mu^+\nu_\mu$ becomes kinematically forbidden for $E_\nu
> 5$ GeV if taking $\delta_{\nu\gamma} =2.5\times 10^{-5}$ as
shown by OPERA data. The detection of $\sim 40$ GeV and $\sim 200$
GeV neutrinos by OPERA and Fermilab can constrain the LIV
parameters to $3\times 10^{-7}$ and $2\times 10^{-8}$
respectively.

Then we discuss if it is possible to constrain the neutrino LIV
parameters by studying the three body decay process of neutrinos.
Such process will modify the atmospheric and astrophysical
neutrino spectra. Because there is no spectral cutoff of the
atmospheric neutrinos up to $\sim 400$ TeV as measured by IceCube
\cite{icecube}, the neutrino LIV parameter can be constrained to
less than $10^{-12}$.

We further studied the astrophysical neutrino spectra from distant
sources, for example the Galactic SNRs, which are usually taken as
the sources of Galactic cosmic rays. The long travel distance
makes it possible to probe even smaller LIV parameters. As an
illustration, we find that for $\xi\sim10^{-15}$ and source
distance $\sim$ kpc, the decay process is expected to result in a
distinctive spectral cutoff of the neutrino spectra at energies
about several tens TeV. This effect can be tested through the
future detection of the neutrino sources.


\begin{acknowledgments}

We thank Prof. X. M. Zhang and Prof. H. B. Hu for helpful discussions.
This work is supported by the Natural Science Foundation of
China under the grant NO. 11075169, and the 973 project under
grant No. 2010CB833000 and by the Chinese Academy of Science
under Grant No. KJCX2-EW-W01.

\end{acknowledgments}

\end{document}